\documentclass[preprint]{aastex}

\slugcomment{Accepted for publication in the ApJL}

\shortauthors{Chung et al.}

\begin{document}

\title{THE EFFECT OF SECOND GENERATION POPULATIONS ON THE INTEGRATED COLORS OF METAL-RICH GLOBULAR CLUSTERS IN EARLY-TYPE GALAXIES}

\author{CHUL CHUNG, SANG-YOON LEE, SUK-JIN YOON, AND YOUNG-WOOK LEE} 
\affil{Department of Astronomy \& Center for Galaxy Evolution Research, Yonsei University, Seoul 120-749, Republic of Korea}

\email{chung@galaxy.yonsei.ac.kr}

\altaffiltext{}{
Department of Astronomy \& Center for Galaxy Evolution Research, Yonsei University, Seoul 120-749, {Republic of} Korea; chung@galaxy.yonsei.ac.kr.}

\begin{abstract}

The mean color of globular clusters (GCs) in early-type galaxies is in general bluer than the integrated color of halo field stars in host galaxies. 
Metal-rich GCs often appear more associated with field stars than metal-poor GCs, yet show bluer colors than their host galaxy light. 
Motivated by the discovery of multiple stellar populations in Milky Way GCs, we present a new scenario in which the presence of second-generation (SG) stars in GCs is responsible for the color discrepancy between metal-rich GCs and field stars. 
The model assumes that the SG populations have enhanced helium abundance as evidenced by observations, and it gives a good explanation of the bluer optical colors of metal-rich GCs than field stars as well as strong Balmer lines and blue UV colors of metal-rich GCs. 
Ours may be complementary to the recent scenario suggesting the difference in stellar mass functions (MFs) as an origin for the GC-to-star color offset. 
A quantitative comparison is given between the SG and MF models.
\end{abstract}


\section{INTRODUCTION}

Globular clusters (GCs) are the oldest stellar systems in galaxies and thus usually considered as the most representative tracer of early star formation in parent galaxies.
Therefore, after the first discovery of bimodal color distribution of GCs in early-type galaxies (ETGs) together with the mean color offset (i.e., mean metallicity offset) between GC systems in ETGs and their host field stars, the color discrepancy between GC systems and their host galaxies was generally interpreted as a different evolutionary history of these two systems \citep{2001MNRAS.322..257F,2002MNRAS.333..383B,2003MNRAS.340..341B, 2005MNRAS.357...56F,2007MNRAS.382.1947F,2007ApJ...665..295P,2011ApJ...728..116L}. 
Recently, \citet{2013ApJ...762..107G} reported that, even though the mean metallicity of red GCs, i.e., metal-rich GCs, and their field stars is almost identical, red GCs in ETGs show bluer colors than integrated colors of field stars in the host galaxies.
In order to interpret this observation, they suggested that star clusters which initially formed with a ``bottom-heavy'' initial mass function (IMF) came to have a ``top-heavy'' IMF due to the long-term dynamical evolution, and their models reproduce the color offset between GCs and field stars of their host galaxies at the same metallicity (hereafter, in short, we will call this as the GC-to-star color offset). 

In this Letter, we suggest an alternative scenario that can also explain the GC-to-star color offset observed in ETGs. 
Our scenario is based on the presence of helium-enhanced stellar populations that can reproduce extremely blue horizontal branch (HB) stars found in the Milky Way GCs (MWGCs) with multiple-generation stellar populations. 
As shown in \citet{2011ApJ...740L..45C}, the hot HB stars from a helium-enhanced subpopulation naturally reproduce the extremely blue integrated $({\rm FUV}-V )_0$ colors of metal-rich GCs in M87 \citep{2006AJ....131..866S, 2007MNRAS.377..987K}, and the effect of helium-enhanced subpopulations on the model $({\rm FUV}-V)_0$ color of metal-rich GCs is far greater than that of metal-poor GCs. 
We extend this result to the optical $(g-z)_0$ and $(B-I)_0$ colors and explain the GC-to-star color offset without applying IMF slope variations in the population models.
  
\section{POPULATION SYNTHESIS MODELS}

The models presented in this Letter are based on the Yonsei Evolutionary Population Synthesis (YEPS).
The detailed prescriptions for the construction of the YEPS model are presented in the two recent papers of \citet{2011ApJ...740L..45C} and \citet{2013ApJS..204....3C}.
We have applied helium-enhanced subpopulations in the models to investigate the effect of second-generation (SG) populations with enhanced helium abundance on metal-rich stellar populations.

\begin{figure}
\includegraphics[angle=-90,scale=0.67]{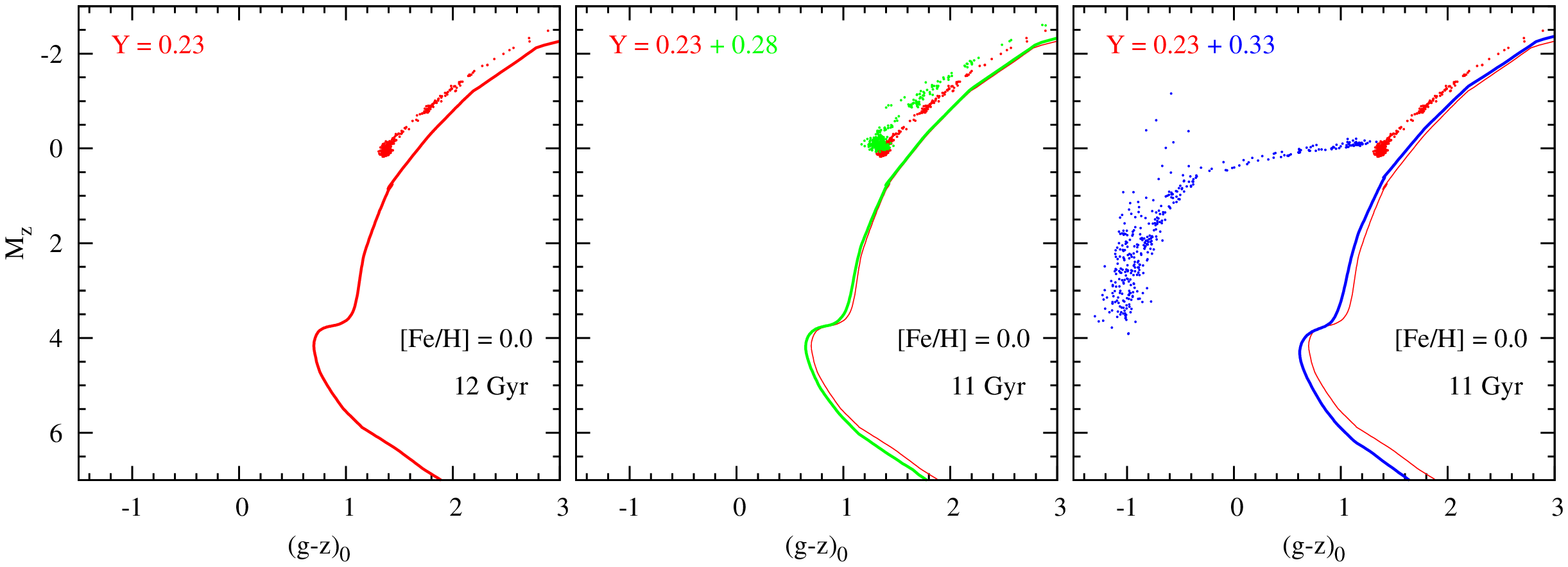}
\caption[]{
Effect of helium abundance on the morphology of HB stars in the color-magnitude diagram of simple stellar populations. 
From left to right panels, the helium abundance of stellar populations in each panel is $Y=0.23$ (red), 0.28 (green), and 0.33 (blue), respectively.
The age of the helium-enhanced SG populations ($Y=0.28$ and 0.33) is 1~Gyr younger than that of the FG population with normal helium abundance ($Y=0.23$).
For the comparison with normal-helium stellar populations, the CMD of the $Y=0.23$ population is overlaid in the middle and right panels.
}
\label{f1}
\end{figure}

Figure~\ref{f1} illustrates the effect of helium-enhanced stellar populations on the morphology of HB stars with [Fe/H]~$= 0.0$.
We assume that the age of the helium-enhanced SG populations in the synthetic color magnitude diagrams (CMDs) is 1~Gyr younger than that of the first-generation (FG) population with $Y=0.23$.
In a single simulation, the number of HB stars in the FG and SG models is approximately 350 and 560, respectively. In addition, in order to avoid small number statistics in HB modeling, we have repeated the simulation 10 times to get the averaged number of HB stars in FG and SG populations. The larger number of HB stars in SG models is caused by their lower mean mass of HB stars that makes their lifetime longer.
Since the mean mass of HB stars with normal helium abundance ($Y=0.23$) is mainly controlled by metallicity and age, metal-rich and relatively young stellar populations show red HB types in the CMDs.
These effects hold for the stellar population with $Y=0.28$ at 11~Gyr.
However, the effect of helium\footnote{We consider the helium content as the third parameter of determining HB types in MWGCs with multiple generation stellar populations, and the difference in helium content among MWGCs is attributed to the different formation efficiency of SG stars with enhanced helium abundance \citep{2008MNRAS.390..693D, 2013ApJ...762...36J}.} on the morphology of HBs overtakes the effect of metallicity and age when the helium abundance reaches $Y=0.33$. 
In the right panel of Figure~\ref{f1}, although metallicity of the stellar population is as high as $[{\rm Fe/H}]=0.0$, the morphology of HBs of the stellar population with $Y=0.33$ shows an extremely blue HB type.
While the color difference of turn-off stars is $< 0.2$~mag between the $Y=0.23$ and 0.33 populations, the mean color difference of HB stars is over $\sim 2$~mag in $(g-z)_0$ color.
This means that the integrated optical colors of metal-rich stellar populations with helium-enhanced SG populations can be substantially bluer than that of models for the FG population with normal-helium abundances only.

Using the SG population with enhanced helium abundance, we have constructed the model for GCs and field stars as follows.
Firstly, we adopt $Y=0.33$ for the helium abundance of the SG population and set the fraction of the SG population for GCs to be 70~\%.
These values are comparable to the simulated mean abundance of helium \citep{2005ApJ...621L..57L, 2013ApJ...762...36J} and the observed fraction of SG populations \citep{2008MNRAS.390..693D} in the MWGCs with multiple stellar populations.
One may argue that the assumed SG fraction in our model is rather too high, but note that the fraction of SG populations in some GCs like M5, M13, M15, M53, and NGC~6397 is estimated to be more than 70~\% based on the analysis of \citet{2008MNRAS.390..693D}. 
Several lines of evidence suggest that 7 -- 8~\% of stars in the inner halo were originated from SG stars in massive GCs \citep{2011A&A...534A.136M}.
Therefore, we have also adopted 10~\% of the SG population with $Y=0.33$ for field stars in host galaxies because the MW halo might be regarded as the local counterpart of ETGs.
Secondly, we assume the age gap between two generations as 1~Gyr\footnote{The significant fraction of ETGs shows recent star formation (RSF) signatures \citep{2011ApJS..195...22Y}. However, the most ETGs with strong bimodal GC color distributions show UV-upturn phenomenon with no signatures of RSF, and the effect of RSF on the integrated colors of ETGs is almost negligible in optical colors (see Fig 4 of \citealt{2011ApJS..195...22Y}). Therefore, we did not include RSF in our simulation for host galaxies.}.
As discussed in \citet{2013ApJ...762...36J}, the age difference between the FG and SG stars for the MWGCs with an enhanced helium abundance is between $0.3 - 1.7$~Gyr.
For all models with SG populations, we have adopted the standard Salpeter IMF, $\xi (m) \propto m^{- \alpha}$, with a slope of $\alpha = 2.35$.
In order to examine the effect of the IMF slope change on the integrated colors of simple stellar populations, we have also simulated models with different IMF slopes of $\alpha = 1.7$ and $3.0$ for FG populations. 
We note that our simulation is designed for both GCs and field stars in ETGs, and thus applying the model to other types of galaxies should be done with caution.
Table~\ref{t1} summarizes the input parameters of models used in this Letter.

\section{RESULTS}

\begin{figure}
\includegraphics[angle=-90,scale=0.61]{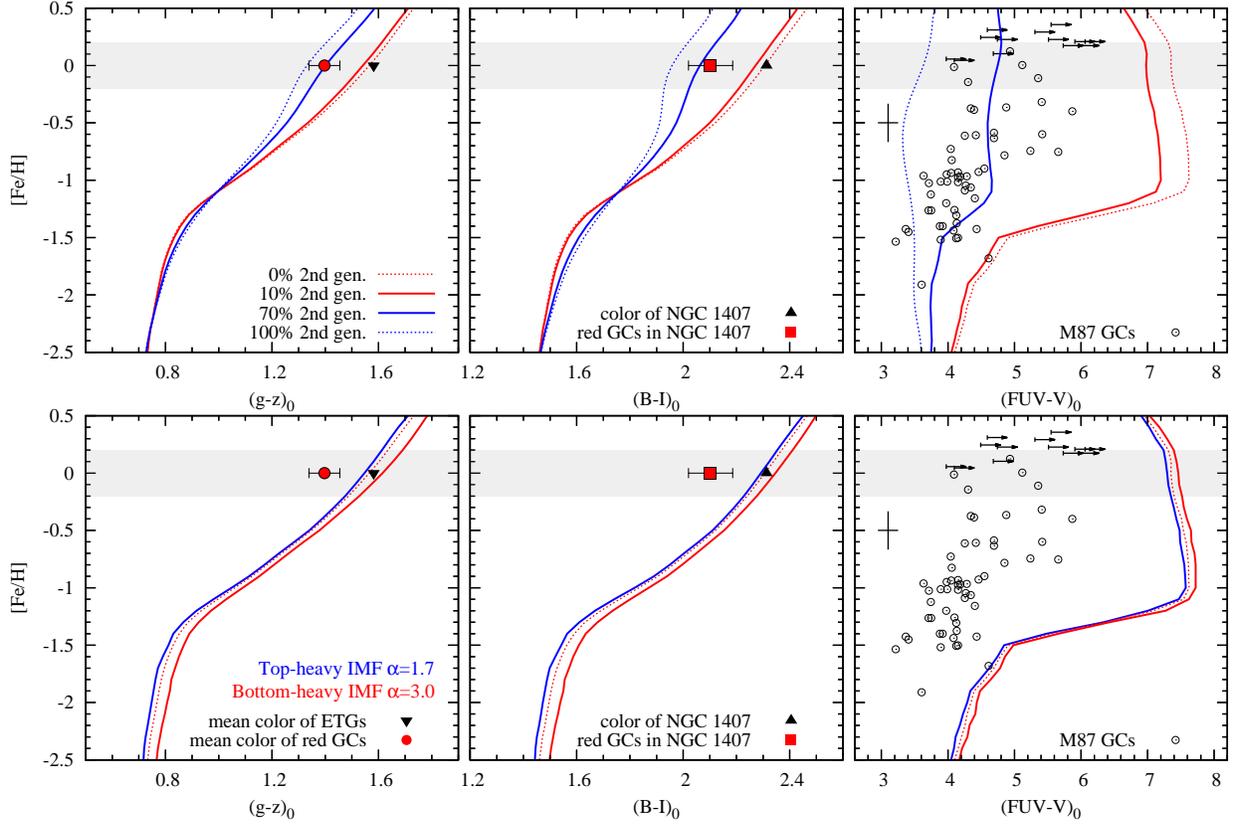}
\caption[]{
Effect of the SG populations on the integrated colors of $(g-z)_0$, $(B-I)_0$, and $({\rm FUV}-V)_0$ ({\it top row}), as well as the effect of the IMF slope on the same integrated colors ({\it bottom row}). 
The age of models in the bottom row is 12~Gyr, while the age of the first and the SG populations in the top row is 12 and 11~Gyr, respectively.
The helium-abundance of the FG and SG populations are $Y=0.23$ and 0.33, respectively.
Red circles and black triangles in the left panels are, respectively, the average color of red GCs and field stars in NGC~4472, NGC~4486, NGC~4649, NGC~4552, NGC~4621, and NGC~4473.
Red squares and black triangles in the middle panels are the colors of red GCs and field stars in NGC~1407.
The GC data in $({\rm FUV}-V)_0$ color are taken from \citet{2006AJ....131..866S}.
Bare arrows in the right panels are upper limits for metal-rich GCs in M87. 
Typical errors for GCs in M87 are indicated.
In the bottom row, the slope $\alpha$ of top-heavy (blue solid lines), bottom-heavy (red solid lines), and the standard IMF model (red dotted lines) is $1.7$, $3.0$, and $2.35$, respectively.
The gray shades in each panel indicate a metal-rich regime from [Fe/H]=$-$0.2 to 0.2.
}
\label{f2}
\end{figure}

Upper panels of Figure~\ref{f2} show the effect of the SG populations on the integrated optical colors of $(g-z)_0$ and $(B-I)_0$, as well as $({\rm FUV}-V )_0$.
For the comparison of models with the observation, we overplot observed colors of red GCs and field stars in the left and middle panels using data in Table~1 of \citet{2013ApJ...762..107G}.
In our simulations, we assume the metallicity of metal-rich GCs as $[{\rm Fe/H}]=0.0$\footnote{Note that if the metallicity of red GCs decreases, the effect of SG populations on the optical colors significantly reduces. Therefore, our simulation presented here can be better applied to metal-rich GCs whose metallicity is greater than $[{\rm Fe/H}]=-0.2$.} because spectroscopy of typical metal-rich GCs suggests the solar metallicity in various observations \citep[e.g.,][]{2003ApJ...592..866C}.
We also presume the observed metallicity of field stars to be [Fe/H]=0.0.
In the metal-rich regime, blue and extremely blue HB stars from helium-enhanced populations make the integrated color of stellar populations bluer. The strongest effect of hot HB stars can be observed in the $({\rm FUV}-V)_0$ color (the right panel in the top row). This effect is not limited to $({\rm FUV}-V)_0$ and makes $(g-z)_0$ and $(B-I)_0$ of the SG population, on average, 0.24 and 0.35~mag bluer than those of the FG population (see the gap between blue and red dotted lines on the gray shades in the top panels). 
The integrated colors of $(g-z)_0$ and $(B-I)_0$ for the metal-rich GC model with the 70~\% of SG populations show, respectively, $0.14$~mag and $0.21$~mag bluer colors than the model for field stars of the host galaxy with the 10~\% of SG populations at $[{\rm Fe/H}]=0.0$. 
Considering the reported GC-to-star color offsets of ETGs ranging from 0.14 to 0.23 for $(g-z)_0$ and 0.21 for $(B-I)_0$ \citep{2013ApJ...762..107G}, our simulation for ${\rm [Fe/H]}=0.0$ well explains GC-to-star color offsets observed in the ETGs without applying the IMF variation.

We have also tested the effect of the IMF variation on the optical colors in the bottom panels of Figure~\ref{f2}.
We assume two different models with the bottom-heavy IMF ($\alpha=3.0$) and the top-heavy IMF ($\alpha=1.7$) compared to the standard Salpeter IMF ($\alpha=2.35$).
The color offsets caused by the variation in the IMF slope from bottom to top heavy, at the same metallicity (${\rm [Fe/H]}=0.0$), are 0.07~mag and 0.06~mag for $(g-z)_0$ and $(B-I)_0$, respectively.
These offsets are not enough to account for the observed GC-to-star offsets in $(g-z)_0$ and $(B-I)_0$. 
Moreover, the model with variations in the IMF hardly reproduces the observed $({\rm FUV}-V)_0$ colors of GCs in M87 (see the right most panels).
It is also noteworthy that the present-day mass function (MF) of clusters deviates significantly from the IMF \citep{2003MNRAS.340..227B, 2010AJ....139..476P}. 
However, the effect of the present-day MF on the integrated color of GCs is very small. 
The color offsets caused by the present-day MF ($\alpha=1.7$) are less than 0.01 mag in both $(g-z)_0$ and $(B-I)_0$ when the [Fe/H] is 0.0.
This result implies that, in order to explain GC-to-star color offsets observed in various colors simultaneously, the inclusion of a SG population in the model is more likely than adopting different IMF slopes in the model. 

\begin{figure}
\includegraphics[angle=-90,scale=0.9]{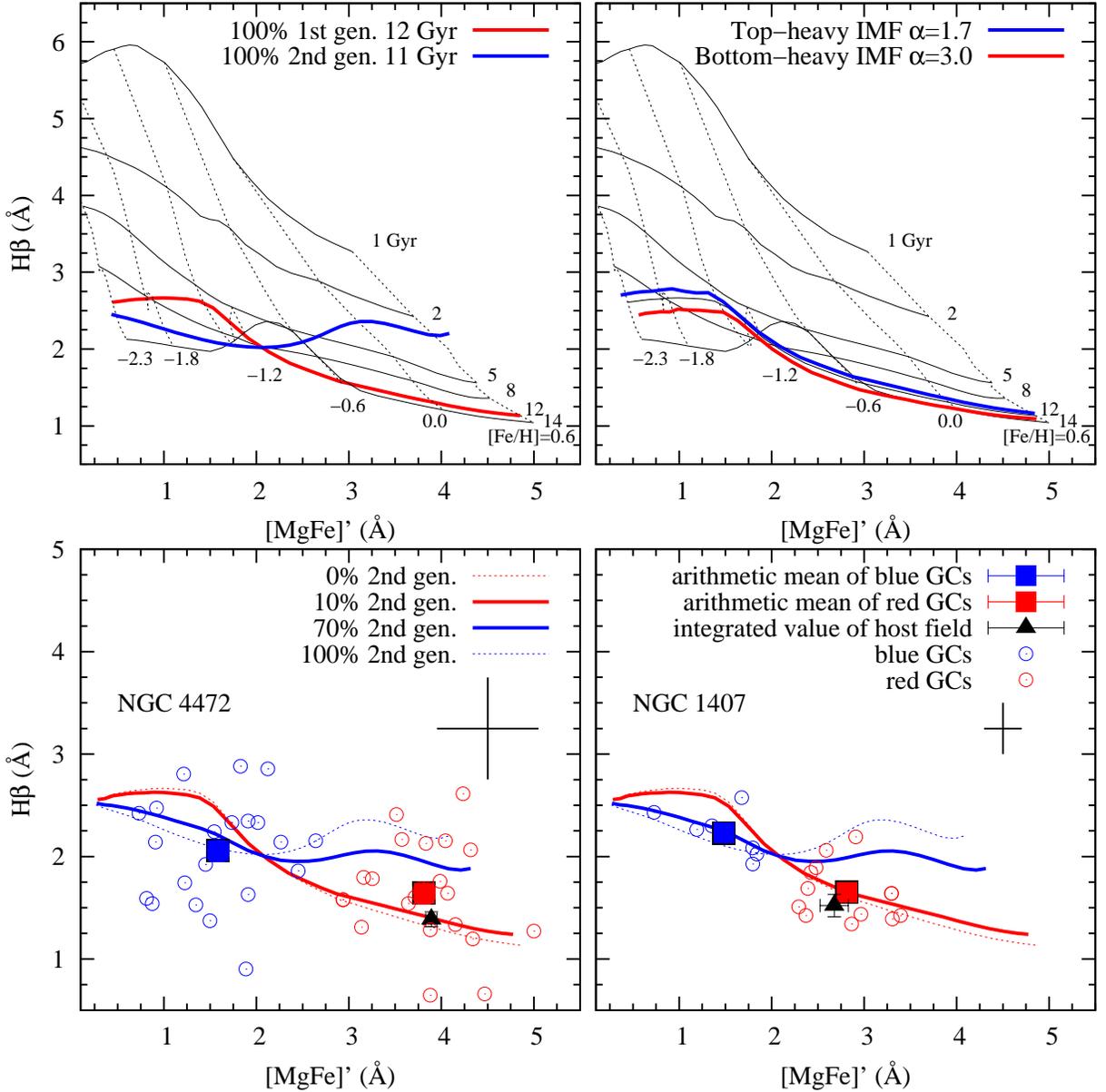}
\caption[]{
({\it top row}) Effects of the SG population and the IMF slope on the integrated absorption indices of the H$\beta$ and the $[{\rm MgFe}]'$ ($\equiv \sqrt{{\rm Mg}b \times (0.72{\rm Fe}5270+0.28{\rm Fe}5335)}$~\AA).
All black lines indicate 100~\% FG models. Red and blue lines in the left panel highlight 100~\% FG and 100\% SG models at the age of 12 and 11 Gyr, respectively.
The red and blue lines in the right panel indicate different IMF slopes with the $\alpha= 1.7$ and $3.0$, respectively.
The age and the metallicity are indicated in the plot.
({\it bottom~row}) Comparison of the models presented in the Figure~\ref{f2} with GCs in NGC~4472 and NGC~1407.
Blue and red open circles are blue and red GCs in NGC~4472 and NGC~1407.
Typical error bars for each GC system are indicated.
Filled blue and red squares indicate arithmetic mean of blue and red GCs in each galaxy NGC~4472 and NGC~1407.
Black triangles are the integrated absorption index of stars in the host galaxies.
}
\label{f3}
\end{figure}

Since the absorption strength of H$\beta$ is very sensitive to the hot temperature ($T_{\rm eff}\sim 10,000$~K) of blue and extremely blue HB stars, it is important to check the effect of the SG population on H$\beta$. 
Figure~\ref{f3} compares our model with observations of NGC~4472 and NGC~1407.
The upper left panel presents the models for FG populations with ages from 1 to 14~Gyr and the model for the 11-Gyr SG population.
Similar to the integrated colors, when [Fe/H] is 0.0, H$\beta$ of the model with the SG population for 11~Gyr shows 1.05~{\AA} stronger absorption than the model of the FG population for 12~Gyr.
This absorption strength is equivalent to the 3~Gyr model of the FG populations.
As shown in the right top panel, the change of the IMF slope from top to bottom-heavy decreases the strength of H$\beta$ for the metal-rich population of $[{\rm Fe/H}]=0.0$ by 0.11~{\AA}.
This is only a 10~\% change with respect to the effect of the SG populations.

In the bottom panels of Figure~\ref{f3}, we compare our models with the GCs in NGC~4472 and NGC~1407 (data are from \citealt{2013ApJ...762..107G}). 
The arithmetic mean of observed H$\beta$ and $[{\rm MgFe}]'$ are also presented. 
Given the observational errors, both models for the FG and the SG population bracket the observed GCs in NGC~4472 and NGC~1407.
The overall GC absorption indices of NGC~4472 and NGC~1407 are better explained by our model with SG populations because the inclusion of the SG populations in the model slightly enhances the strength of H$\beta$ by 0.4~${\rm \AA}$ when the mean [Fe/H] of the metal-rich GCs is 0.0. 
The H$\beta$ difference between two observed quantities of NGC~4472 differs at $\sim 1.96 \sigma$ confidence level. Although the confidence level of the H$\beta$ difference in NGC~1407 is lower ($\sim 1.02 \sigma$), this is not inconsistent with the result of NGC~4472.
In general, the age-dating of GCs in ETGs based on the model only with FG populations gives systematically younger ages for metal-rich GCs because of the scattered distribution of metal-rich GCs.
For example, the age of GCs in NGC~4472 \citep{2003ApJ...592..866C} and NGC~5128 \citep{2010ApJ...708.1335W} become younger with increasing metallicity.
However, the inclusion of a SG population in the model reduces the age-metallicity gradient and reproduce simultaneously integrated optical colors and absorption indices of GCs in ETGs without any further fine tuning.
This result reinforces that the helium-enhanced SG population is a very plausible candidate for the subpopulation in the metal-rich GCs in ETGs.

\section{DISCUSSION}

We have reproduced the observed GC-to-star color offsets found in massive ETGs using helium-enhanced SG stellar populations.
Our models presented here suggest that some 70~\% of helium-enhanced SG population explains GC-to-star color offsets observed in the ETGs without applying the variation in the IMF slope.
In addition, the general trend between H$\beta$ and metallicity found in metal-rich GCs can also be explained by our models with SG populations. 
These results are consistent with our previous models for metal-rich GCs in M87 \citep{2011ApJ...740L..45C}, which showed that the inclusion of helium-enhanced subpopulations in the models can provide the most reasonable explanation for the extremely blue $({\rm FUV}-V)_0$ color.

Our result may shed some light on the long-standing problem of the disagreement between GCs and filed stars in their metallicities.
The discrepancy in metallicities are twofold: (a) the metallicity distribution function (MDF) morphologies and (b) the mean metallicity values. 
\citet{2011ApJ...743..150Y} delved into the difference in MDF morphology and, based on their theoretical color-to-metallicity conversion relations, suggested that inferred GC MDFs are remarkably similar to the MDFs of resolved field stars in nearby ETGs. 
As for the mean metallicity values of GCs and stars, however, the discrepancy remains in the sense that GCs are bluer than field stars within the same galaxy. 
\citet{2011ApJ...743..150Y} attributed the remaining GC-to-star metallicity offset to the difference in their details of chemical evolution processes, proposing a scenario in which GCs are the relics of vigorous star formation early on, and thus trace selectively the major mode of star formation. 
That is, GC formation was less prolonged than field-star formation. As a consequence, the chemical enrichment of GC systems has ceased earlier than that of the field stars, leaving GCs that are metal-poorer than field stars. 
Our SG scenario predicts that metallicity of metal-rich GCs is the same as that of field stars, further reducing the degree of the metallicity difference between GCs and stars. 
Our results, combined with those of \citet{2011ApJ...743..150Y}, indicate that GC systems and their parent galaxies have shared a more common evolutionary history than previously believed.

Since the investigation by \citet{2010Natur.468..940V}, the idea of a bottom-heavy IMF for the massive ETGs has received much attention in the current literature \citep[e.g.,][]{2012Natur.484..485C, 2012ApJ...760...70V, 2012ApJ...760...71C, 2013arXiv1304.1178Z}.
Their interpretations are based on the strengths of the Na~I doublet at $\lambda$ 8183, 8195~{\AA} and the Wing-Ford band at $\lambda$ 9916~{\AA} that become stronger when low mass stars are abundant.
However, it is well known that the strength of spectroscopic absorption line is controlled not only by the gravity but also by the temperature and chemical composition.
Therefore, if field stars in ETGs are contaminated by the SG populations from massive GCs, the strong Na~I doublet of ETGs may also be caused by SG populations that generally contain enhanced Na \citep{2004ARA&A..42..385G, 2005A&A...433..597C, 2009A&A...505..117C}.
The Figure~2b of \citet{2010Natur.468..940V} gives some hints for the effect of the enhanced Na element because their result based on Na~I doublet prefers steeper IMF slope (bottom-heavy IMF) than that based on the Wing-Ford band measurement.
In addition,the Salpeter-like IMF of ETGs are reported based on spectroscopy \citep{2012MNRAS.426.2994S} and stellar dynamics \citep{2011MNRAS.415..545T} of ETGs.
We will investigate this issue in a separate paper.

\acknowledgments

We thank the referee for a number of helpful suggestions.
YWL and SJY acknowledge support from the National Research Foundation of Korea to the Center for Galaxy Evolution Research.
SJY acknowledges support from Mid-career Researcher Program (No. 2012R1A2A2A01043870) through the National Research Foundation (NRF) of Korea grant funded by the Ministry of Education, Science and Technology (MEST), and support by the Korea Astronomy and Space Science Institute Research Fund 2012 and 2013.
This work was partially supported by the KASI-Yonsei Joint Research Program (2012-2013) for the Frontiers of Astronomy and Space Science.

\clearpage

\clearpage

\clearpage

\clearpage

\clearpage

\begin{deluxetable}{lccc}
\tabletypesize{\scriptsize}
\tablewidth{0pt}
\tablecaption{\label{tab:table1} INPUT PARAMETERS ADOPTED IN THE SECOND-GENERATION AND COMPARISON MODELS}
\tablehead{\colhead{Parameters} &\colhead{Models for GCs} &\colhead{Models for host galaxy field stars} &\colhead{Comparison models}}
\startdata
Slope of Salpeter initial mass function, $\alpha$ &$2.35$&$2.35$&$1.7$ and $3.0$ \\
$\alpha$-elements enhancement, [$\alpha$/Fe]& 0.3& 0.3& 0.3 \\
HB mass dispersion, ${{\sigma_{}}_{M}}$ ($M_{\odot}$)&0.015&0.015&0.015\\
Reimers' mass-loss parameter, $\eta$&0.63&0.63&0.63\\
Absolute age, $t$ (Gyr)&11.0 and {12.0}&11.0 and {12.0}& {12.0}\\
Helium enhanced population ratio, $Y$ (\%) &0.23~({30\%})~+~0.33~({70\%})&0.23~(90\%)~+~0.33~(10\%)&0.23 (100\%)\\

\enddata
\label{t1}
\end{deluxetable}

\end{document}